\documentclass[aps,pra,superscriptaddress,amsmath,amssymb,preprintnumbers,twocolumn, showpacs]{revtex4}
\usepackage{amssymb}
\usepackage{color}
\usepackage{graphicx}

\begin{document}
\def\bbm[#1]{\mbox{\boldmath$#1$}}

\title{Reconstruction of Markovian Master Equation parameters through symplectic tomography.}

\author{Bruno Bellomo}\affiliation{MECENAS, Universit\`{a} Federico II di Napoli, Via
Mezzocannone 8, I-80134 Napoli, Italy}\affiliation{CNISM and
Dipartimento di Scienze Fisiche ed Astronomiche, Universit\`a di
Palermo, via Archirafi 36, 90123 Palermo, Italy}

\author{Antonella De Pasquale}\affiliation{MECENAS, Universit\`{a} Federico II di Napoli, Via
Mezzocannone 8, I-80134 Napoli, Italy}\affiliation{Dipartimento di
Fisica, Universit\`a di Bari,
        I-70126  Bari, Italy;\\INFN, Sezione di Bari, I-70126 Bari,
        Italy}

\author{Giulia Gualdi}\affiliation{MECENAS, Universit\`{a} Federico II di Napoli, Via
Mezzocannone 8, I-80134 Napoli, Italy}\affiliation{Dipartimento di Fisica, Universit\`a di Camerino, I-62032 Camerino (MC), Italy}

\author{Ugo Marzolino}\affiliation{MECENAS, Universit\`{a} Federico II di Napoli, Via
Mezzocannone 8, I-80134 Napoli, Italy}
\affiliation{Dipartimento di Fisica Teorica, Universit\`a di Trieste, Strada Costiera 11, 34014 Trieste,
Italy; \\
INFN, Sezione di Trieste, 34014 Trieste, Italy}

\begin{abstract}

In open quantum systems, phenomenological master equations with
unknown parameters are often introduced. Here we propose a
time-independent procedure based on quantum tomography to
reconstruct the potentially unknown parameters of a wide class of
Markovian master equations. According to our scheme, the system
under investigation is initially prepared in a Gaussian state.  At
an arbitrary time $t$, in order to retrieve the unknown coefficients
one needs to measure only a finite number (ten at maximum)
of points along three time-independent tomograms. Due to the
limited amount of measurements required, we expect our
proposal to be especially suitable for experimental implementations.
\end{abstract}


\maketitle

\section{Introduction}

Tomographic maps \cite{Asorey2007} can be considered a very useful
tool for reconstructing the physical state or some other properties
of many physical systems, both in a classical
(e.g. medical physics, archaeology, biology, geophysics) and in a quantum
perspective (e.g. photonic states \cite{Smithey1993}, photon number distributions
\cite{Brida2006, Zambra2005, Zambra2006}, longitudinal motion of
neutron wave packets \cite{Badurek2006}).

The tomographic analysis is based on a probabilistic approach towards  physical system
investigation. In particular, its key ingredient  is
 the Radon transform \cite{Radon1917}.  Given the phase-space of the system,
 this invertible integral transform allows to retrieve the marginal probability densities of the system, i.e. the probability density along straight lines.
However, while in the classical regime the  state of the system can
be fully described by means of a probability distribution on its
phase space, this is no longer the case of quantum
systems.
Indeed, due to the Heisenberg uncertainty relation,
it is not possible to write a probability distribution as a function
of  both momentum and position.  In this case, the Wigner function
\cite{Wigner1932, Moyal949} can be employed as a quantum
generalization of a classical probability distribution. This
function is  a map between
 phase-space  functions  and density matrices.
Even if the Wigner function can take on negative values, by
integrating out either the position or the momentum  degrees of
freedom, one obtains a bona fide probability distribution
for the conjugated variables. From this point of view, the Wigner
function corresponding to a quantum state can be regarded as a
quasi-probability distribution and interpreted as a joint
probability density in the phase space \cite{Man'ko1999}.

In this paper we apply quantum symplectic tomography to the investigation of
open quantum systems
\cite{Petruccione-Breuerlibro2002,BenattiFloreanini2005}
which, due to the  coupling to an environment (bath),  undergo a non-unitary dynamical evolution. A complete microscopic description of system-plus-bath dynamics is a complex many-body problem. Hence, as in general  one aims at describing the dynamics of the system, only basic information about the bath is retained, according to the so-called open system approach.
The state of the system is then expressed by means of a reduced density matrix, obtained  from the total density matrix by tracing out the environmental degrees of freedom. The system dynamics is then governed by the so-called quantum master equation. The master equation approach
 can be seen as the generalization of the Schr\"odinger equation to the possibly  incoherent evolution of a density matrix.  In this case, the generator of the time evolution is  the Liouville dissipative operator.
The integration of a time-dependent Liouvillian being a highly involved task, e. g. see \cite{HPZ1992,Halliwell1996}, it is highly preferrable  to deal with a time-independent Liouvillian, i.e. to assume a Markovian dynamics. Several approximations allow a Markovian description, such as  the weak coupling limit, the singular limit and
the low density limit
\cite{Petruccione-Breuerlibro2002,BenattiFloreanini2005,spohn1980}.\\
Nevertheless, a proper  derivation of the master equation still
requires complete information about the bath. The lack of this
knowledge leads to the derivation of phenomenological master
equations with unknown coefficients. Indeed, recent investigations
\cite{Lidar2001,SchallerBrandes2008,BFM2009} provide a more accurate
approximation than the weak coupling limit, due to a more
refined coarse grained dynamics. Even in this case, the
obtained master equation has unknown coefficients, as it depends
phenomenologically on the system investigated.

In this paper, we will focus on a class of Markovian master equations with unknown coefficients modeling a one-dimensional damped harmonic oscillator.
 In particular,  we choose  Lindblad operators  \cite{Lindblad1976} linear in both momentum and position degrees of freedom, such that the dynamical evolution of the system preserves  the  Gaussian form of the states.
Our goal is to show how,  by means of a tomographic approach, it is possible to measure indirectly the unknown coefficients by using Gaussian wave packets as a probe.

This paper is organized as follows. In section \ref{par:Description
of the system} we introduce the class of master equations we want to
investigate.  In section \ref{Evolution of gaussian wave packet} we
derive the expressions for the coefficients of the master equation
as a function of the first and second evolved momenta (cumulants) of
a Gaussian state. In section \ref{Par:Reconstruction of the
cumulants of a gaussian state with a tomographic approach} we
introduce the Wigner function and the Radon transform for an
arbitrary Gaussian wave packet at a generic time $t$ . We show that
in order to measure the cumulants of a Gaussian state, and then
indirectly the unknown parameters of the master equation, we need
only a finite number (eight or ten) of time-independent tomograms.
 In section
\ref{par:Summary and Conclusions} we summarize and discuss our
results and outline some feasible applications.
Finally, in appendix A, we propose an alternative procedure
to obtain the cumulants of a Gaussian state by means of
time-dependent tomograms. This approach  however appears to be less
convenient for practical implementations.

\section{Description of the system}\label{par:Description of the system}
We want to investigate a class of master equations describing a
Gaussian-shape-preserving (GSP) evolution of a quantum state. In the
Markovian approximation,  the non-unitary time evolution of a
quantum system is described by the following general master equation
\cite{Lindblad1976,GKS1976}:
\begin{eqnarray}\label{lindblad form}
    \frac{d \hat{\rho}(t)}{d t}=&&L(\hat{\rho}(t))\nonumber\\=&-&\frac{i}{\hslash}\left[\hat{H}, \hat{\rho}(t) \right]\nonumber\\&+&\frac{1}{2\hslash}\sum_j
\left(\left[\hat{V}_j \hat{\rho}(t),\hat{V}_j^\dag
\right]+\left[\hat{V}_j \hat{\rho}(t)\hat{V}_j^\dag\right]\right),\label{hatro}
\end{eqnarray} where $\hat\rho(t)$ is the reduced density operator of the system.
Eq. (\ref{hatro}) is exactly solvable if the Lindblad operators $\hat{V}_j$ and the system Hamiltonian $\hat{H}$ are, respectively,  at most first and second degree polynomials in position  ($\hat{q}$) and momentum ($\hat{p}$) coordinates \cite{Sandulescu1987,Isar1994}.

For systems like  a harmonic oscillator or a
field mode in an environment of harmonic oscillators (i.e.
collective modes or a squeezed bath), $\hat{H}$ can
be chosen of the general quadratic form
\begin{equation}\label{Hamiltonian}
    \hat{H}=\hat{H}_0+\frac{\delta}{2}\left(\hat{q}\hat{p}+\hat{p}\hat{q}\right), \qquad \hat{H}_0= \frac{1}{2m}\hat{p}^2+\frac{m\omega^2}{2}\hat{q}^2,
\end{equation}
where $\delta$ is the strength of the bilinear term in $\hat{q}$ and
$\hat{p}$, $m$ is oscillator mass, and $\omega$ its frequency. The
operators $\hat{V}_j$, which model the environment, are  linear
polynomials in $\hat{q}$ and $\hat{p}$:
\begin{equation}\label{linbblad operators}
    \hat{V}_j=a_j \hat{p} + b_j \hat{q}, \quad j=1,2,
\end{equation}
with $a_j$ and $b_j$ complex numbers. The sum goes from $1$ to $2$
as  there exist only two c-linear independent operators
$\hat{V}_1$, $\hat{V}_2$, in the linear space of first degree polynomials in $\hat{p}$ and $\hat{q}$.
We can safely omit generic constant contributions in
$\hat{V}_j$ as they do not influence the dyamics of the system.

Given this choice of operators, the Markovian master equation (\ref{lindblad form}) can be rewritten as:
\begin{eqnarray}\label{master equation in q e p}
\frac{\mathrm{d} \hat{\rho}(t)}{\mathrm{d}
t}&=&\!-\frac{i}{\hslash}\left[\hat{H}_0,\hat{\rho}(t)
\right]-\frac{i (\lambda +\delta)}{2
\hslash}\left[\hat{q},\hat{\rho} (t) \hat{p}+\hat{p}\hat{\rho}(t) \right]
\nonumber\\&& +\frac{i (\lambda -\delta)}{2 \hslash}\left[\hat{p},\hat{\rho}(t)
\hat{q}+\hat{q}\hat{\rho}(t) \right]\nonumber\\&&-\frac{D_{pp}}{
\hslash^{2}}\left[\hat{q},[\hat{q},\hat{\rho}] \right]
-\frac{D_{qq}}{ \hslash^{2}}\left[\hat{p},[\hat{p},\hat{\rho}(t)]
\right]\nonumber\\&&+\frac{D_{qp}}{ \hslash^{2}}
\left(\left[\hat{q},[\hat{p},\hat{\rho}(t)]\right]+\left[\hat{p},[\hat{q},\hat{\rho}(t)]
\right]\right)\label{mark}
\end{eqnarray}
where $\lambda=-\mathrm{Im}\sum_{j=1,2}a_j^*b_j$ is the unknown friction
constant and
\begin{eqnarray}\label{D coefficients}
    D_{qq}&=&\frac{\hslash}{2}\sum_{j=1,2}|a_j|^2, \quad D_{pp}=\frac{\hslash}{2}\sum_{j=1,2}|b_j|^2,\nonumber\\
    D_{qp}&=&-\frac{\hslash}{2}\mathrm{Re}\sum_{j=1,2}a_j^*b_j\end{eqnarray}
are the unknown diffusion coefficients, satisfying  the following
constraints which ensure the complete positivity of the time evolution \cite{Sandulescu1987,Isar1994}:
\begin{equation}\label{constraints}
    i) D_{qq}>0,\quad ii) D_{pp}>0,\quad iii) D_{qq}D_{pp}-D_{qp}^2\geqslant \lambda^2\hslash^2/4.
\end{equation}

Markovian GSP  Master equations of the form Eq.~(\ref{master
equation in q e p}) are used  in quantum optics and nuclear physics
\cite{Savage1985,Kennedy1988,Yang1986}, and in the limit of
vanishing $\omega$ can be employed for a phenomenological
description of quantum Brownian motion
\cite{Caldeira1983,Barnett2005, bellomo2007a}. Also, in the case of
a high-temperature Ohmic environment  the time-dependent master
equation derived in \cite{HPZ1992,Halliwell1996} can be recast in
this time-independent shape. It must be noted however that  in the
high-temperature limit  the third constrain in  (\ref{constraints})
seems to be violated.  Nevertheless, even if $D_{qq}=0$, $D_{qp}=0$
and $\lambda\neq 0$, $D_{pp}$ diverges only  linearly with
temperature. Therefore, we can recover the complete positivity by
means of a suitable renormalization. This renormalization consists
in adding a suitable subleading term $D_{qq}$ (e.g. $D_{qq}\propto
T^{-1}$). Otherwise, we can consider an high frequency cut-off for
the environment \cite{HPZ1992,Halliwell1996}. In this way the
master equation is not Markovian anymore. Anyway, since it involves only
regular functions, it should give a completely positive dynamics (as
the microscopic unitary group does).

\section{Gaussian states evolution}\label{Evolution of gaussian wave packet}
In this section we investigate the evolution  of an initial Gaussian state according to Eq. (\ref{master equation in q e p}). In particular we derive invertible expressions for  the cumulants of the state at a time $t$ in terms of the parameters of the master equation.
Due to the Gaussian shape preservation, the evolved state at time $t$ is completely determined by its first and second order momenta:
\begin{eqnarray}\label{cumulants}
\langle \hat{q}\rangle_t &=& {\rm Tr}(\hat{\rho}(t)\hat q), \nonumber \\
\langle \hat{p}\rangle_t &=& {\rm Tr}(\hat{\rho}(t)\hat p), \nonumber \\
\Delta q_t^2 &=& {\rm Tr}(\hat{\rho}(t)\hat q^2)-\langle \hat{q}\rangle^2_t, \nonumber \\
\Delta p_t^2 &= & {\rm Tr}(\hat{\rho}(t)\hat p^2)-\langle \hat{p}\rangle^2_t, \nonumber \\
\sigma(q,p)_t & = & {\rm Tr}\left(\hat{\rho}(t)\frac{\hat{q}\hat{p}+\hat{p}\hat{q}}{2}\right)-\langle \hat{q}\rangle_t\langle \hat{p}\rangle_t.
\end{eqnarray}
Due to the linearity of the $\hat{V}_j$'s in phase-space, the
time-evolution of the first and second order cumulants can be
decoupled.
We then obtain the following  two sets of solvable
equations   \cite{Sandulescu1987,Isar1994}:
\begin{equation}
\left\{
\begin{array}{rl}
\frac{d}{dt}\langle\hat{q}\rangle_t&=-(\lambda-\delta)\langle\hat{q}\rangle_t+\frac{1}{m}\langle\hat{p}\rangle_t \\
\\\frac{d}{dt}\langle\hat{p}\rangle_t&=-m\omega^2\langle\hat{q}\rangle_t-(\lambda+\delta)\langle\hat{p}\rangle_t
\end{array}\right.\label{gaussevol1}
\end{equation}
\begin{equation}
\left\{
\begin{array}{rl}
\frac{d}{dt}\Delta q_t^2 & \displaystyle
=-2(\lambda-\delta)\Delta q_t^2+\frac{2}{m}
\sigma(q,q)_t+2D_{qq} \\ \\
\frac{d}{dt}\Delta p_t^2 & \displaystyle =-2(\lambda+\delta)\Delta p_t^2-2m\omega^2\sigma(q,p)_t+2D_{pp} \\
\frac{d}{dt}\sigma(q,p)_t & \displaystyle =-m\omega^2\Delta
q_t^2+\frac{1}{m}\Delta p_t^2 -2\lambda\sigma(q,p)_t+2D_{qp}
\end{array}\right.
\label{gaussevol2}\end{equation}
The above
equations allow to obtain the time-dependent momenta as a function of the Master equation coefficients
$\lambda,D_{qq},D_{pp},D_{qp}$.
We now show how to invert these relations in order to express
 the parameters $\lambda,D_{qq},D_{pp},D_{qp}$ as a function of
the evolved cumulants at an arbitrary time.
The solution of Eqs. (\ref{gaussevol1}) is given by
\cite{Sandulescu1987,Isar1994}
\begin{equation} \label{evolved averages}
\begin{cases}
\langle\hat q\rangle_t= & \displaystyle e^{-\lambda t}\Bigg[\langle\hat q\rangle_0\left(\cosh\eta t+\frac{\delta}{\eta}\sinh\eta t\right)\\
& \displaystyle \qquad+\langle\hat p\rangle_0\frac{1}{m\eta}\sinh\eta t\Bigg] \\
\langle\hat p\rangle_t= & \displaystyle e^{-\lambda
t}\Bigg[-\langle\hat q\rangle_0\frac{m\omega^2}{\eta}\sinh\eta
t\\
& \displaystyle \qquad+\langle\hat p\rangle_0\left(\cosh\eta
t-\frac{\delta}{\eta}\sinh\eta t\right)\Bigg],
\end{cases}
\end{equation}
where $\eta^2=\delta^2-\omega^2$. If $\eta^2<0$ we can set
$\eta=i\Omega$ and the previous equations hold again with
trigonometric instead of hyperbolic functions. The
coefficient $\lambda$ can then be obained by inverting
 Eqs.\,(\ref{evolved averages}). \\
The elements of the diffusion matrix can be retrieved from the
second set of equations (\ref{gaussevol2}), whose solutions can be
expressed in a compact form as
\begin{equation}
X(t)=(Te^{Kt}T)X(0)+T K^{-1}(e^{Kt}-1)TD,\label{evolv}
\end{equation}
where

\begin{eqnarray}
X(t)&=&
\begin{pmatrix}
\displaystyle m\omega\Delta q_t^2 \\
\displaystyle \frac{\Delta p_t^2}{m\omega} \\
\displaystyle \sigma(q,p)_t
\end{pmatrix},
\qquad D=
\begin{pmatrix}
\displaystyle 2m\omega D_{qq} \\
\displaystyle \frac{2D_{pp}}{m\omega} \\
\displaystyle 2D_{qp} \\
\end{pmatrix}, \nonumber \\
T&=&\frac{1}{2\eta}
\begin{pmatrix}
\delta+\eta & \delta-\eta & 2\omega \\
\delta-\eta & \delta+\eta & 2\omega \\
-\omega & -\omega & -2\delta \qquad\end{pmatrix}, \nonumber
\\K&=&
\begin{pmatrix}
-2(\lambda-\eta) & 0 & 0 \\
0 & -2(\lambda+\eta) & 0 \\
0 & 0 & -2\lambda \qquad\end{pmatrix}.
\end{eqnarray}
>From the invertibility of matrices $T$ ($T^2=1$) and
$\tilde{K}=K^{-1}\left(e^{Kt}-1\right)$ (invertible for bounded $K$
also if some of its eigenvalues are 0), we can  derive the
expression of $D_{qq}$, $D_{pp}$ and $D_{qp}$ using Eq.
(\ref{evolv}):

\begin{eqnarray}
D&=&T\tilde K^{-1}T\left(X(t)-(Te^{Kt}T)X(0)\right), \nonumber\\ \\
\tilde K&=&K^{-1}(e^{Kt}-1)\nonumber\\&=&
\begin{pmatrix}
\displaystyle \frac{1-e^{-2(\lambda-\eta)}}{2(\lambda-\eta)} & 0 & 0 \\
0 & \displaystyle \frac{1-e^{-2(\lambda+\eta)}}{2(\lambda+\eta)} & 0 \\
0 & 0 & \displaystyle \frac{1-e^{-2\lambda}}{2\lambda}
\end{pmatrix}.\quad
\end{eqnarray}

We emphasize that the time $t$ at which we are considering the
cumulants is completely arbitrary. For instance, the expression of
the coefficients $D_{qq},D_{pp},D_{qp}$ in terms of the asymptotic
second cumulants and the parameter $\lambda$ reads:
\begin{eqnarray}
D_{qq} & = & (\lambda-\delta)\Delta q_\infty^2-\frac{1}{m}\sigma(q,p)_\infty, \nonumber \\
D_{pp} & = & (\lambda+\delta)\Delta p_\infty^2+m\omega^2\sigma(q,p)_\infty, \nonumber \\
D_{qp} & = &
\frac{1}{2}\left(m\omega^2\Delta q_\infty^2-\frac{1}{m}\Delta p_\infty^2+
2\lambda\sigma(q,p)_\infty\right).\nonumber \\
\end{eqnarray}

\section{Cumulants reconstruction through tomography}\label{Par:Reconstruction of the cumulants of a gaussian state with a tomographic approach}

In this section we introduce a procedure based on
symplectic tomography in order to measure the first and second
cumulants of a Gaussian wave packet at an arbitrary time $t$.
This will allow us to indirectly measure
the parameters $\lambda,D_{qq},D_{pp},D_{qp}$, them being  functions of
the evolved cumulants at an arbitrary time (see previous section).
The tomographic approach is very useful when dealing with
a phenomenological master equation of the form of Eq.~(\ref{master equation in q e
p}), as
the dependence of the coefficients of the master equation from
the physical parameters is in principle unknown.

\subsection{Symplectic tomography}\label{par:tomography}

Given a quantum state $\hat{\rho}(t)$ its Wigner function reads:
\begin{equation}\label{wigner function definition}
    W(q,p,t)=\frac{1}{\pi \hslash} \int_{-\infty}^{+\infty}\mathrm{d}y \exp \left(
    \frac{i 2 p y}{\hslash}\right)\hat{\rho} (q-y,q+y,t). \\
\end{equation}\\
If the system dynamics is described by the master equation
 (\ref{master equation in q e p}),  and the initial state is Gaussian,
 the Wigner function preserves the Gaussian form of the state.
Indeed, it can be expressed  as a function of its first and second order
momenta:
\begin{eqnarray}\label{wigner function gaussian}
&& W(q,p,t)  =  \frac{1}{2 \pi \sqrt{\Delta q_t^2\Delta p_t^2-\sigma(q,p)_t^2}}
  \nonumber\\  &&\times \exp \Bigg[-  \frac{\Delta q_t^2
  (p-\langle \hat{p}\rangle_t)^2+\Delta p_t^2(q-\langle \hat{q}\rangle_t)^2}{2[\Delta q_t^2\Delta p_t^2-\sigma(q,p)_t^2]}\nonumber\\&&\qquad\quad-\frac{2\sigma(q,p)_t (q-\langle \hat{q}\rangle_t)(p-\langle \hat{p}\rangle_t)}{2[\Delta q_t^2\Delta p_t^2-\sigma(q,p)_t^2]} \Bigg]
  \,.
\end{eqnarray}
Let us now consider the line in phase-space
\begin{equation}\label{line in q p plane}
    X-\mu q - \nu p = 0.
\end{equation}
The tomographic map of a generic state
along this line, i.e. its Radon transform, is given by:
\begin{eqnarray}\label{radon transform}
    \varpi (X,\mu,\nu)&=&\langle \delta\left(X-\mu q - \nu p\right)
    \rangle\nonumber\\ &=&
    \int_{\mathbb{R}^2} W(q,p,t)  \delta\left(X-\mu q - \nu p\right) \mathrm{d}q
    \mathrm{d}p.\nonumber\\
\end{eqnarray}
>From equation \,(\ref{wigner function gaussian}) it follows that for
a Gaussian wave packet  the
Radon transform can be explicitly written as:
\begin{multline}\label{radon transform gaussian}
    \varpi (X,\mu,\nu)=
    \frac{1 }{\sqrt{2\pi}\sqrt{\Delta q_t^2\mu^2+\Delta p_t^2 \nu^2 + 2\sigma(q,p)_t \mu \nu}} \\\exp \left[-\frac{\left(X-\mu \langle \hat{q}\rangle_t - \nu \langle \hat{p}\rangle_t \right)^2}{2[\Delta q_t^2\mu^2+\Delta p_t^2 \nu^2 + 2\sigma(q,p)_t \mu \nu]}
    \right]\,,
\end{multline}
with  the following constraint on the second cumulants:
\begin{equation}
\Delta q_t^2\mu^2+\Delta p_t^2 \nu^2 + 2\sigma(q,p)_t \mu \nu
>0.\end{equation} This constrain is obeyed for each value of the
parameters $\mu$ and $\nu$ iff $\Delta q_t^2\Delta
p_t^2-\sigma(q,p)_t^2>0$. This inequality is always satisfied as
a consequence of the Robertson-Schr\"{o}dinger relation.

 Eq.~(\ref{radon transform}) also implies a homogeneity
condition on the tomographic map: $|c|\,\varpi (cX,c\mu,c\nu)=\varpi
(X,\mu,\nu)$. This condition can be used in the choice of
parameters $\mu,\nu$. In fact, if one uses  polar coordinates $(r,
\theta)$, i.e. $\mu=r\cos\theta$, $\nu=r\sin\theta$, the homogeneity
condition can be used to eliminate the parameter $r$. From Eq.
(\ref{line in q p plane}) it emerges that the  coordinates of the
phase space need to be properly rescaled in order to have the same
dimensions.  For instance, we can set $q\to\sqrt{\frac{m
\omega}{\hslash}}q$ and $p\to\sqrt{\frac{1}{\hslash m \omega}}p$. In
particular if $\omega=0$, i.e.  for a free particle
interacting with the environment, we can choose the same
rescaling with a fictitious frequency defined by
$\hslash\bar\omega=\Delta p_0^2/2m$, imposing $q\to\frac{\Delta
p_0}{\sqrt{2}\hslash}q$ and $p\to\frac{1}{\sqrt{2}\Delta p_0}p$.
In general, every rescaling assigning the same dimensions to $q$ and
$p$ is suitable for our purpose.

\subsection{From tomograms to cumulants}\label{From
tomograms to cumulants}

Let us  now consider the tomograms corresponding to two different
directions in phase space, i.e. to two different
 couples of parameters $(\mu, \nu)$, e. g.  $X=q$ and $X=p$. These
lines in phase space are associated respectively to the position and
momentum probability distribution functions:
\begin{eqnarray}
\label{pdfq}
\varpi(X,1,0) & = & \frac{1}{\Delta q_t\sqrt{2\pi}}\exp \left[-\frac{\left(X-\langle \hat{q}\rangle_t\right)^2}{2\Delta q_t^2}\right],  \\
\label{pdfp}
\varpi(X,0,1) & = & \frac{1}{\Delta p_t\sqrt{2\pi}}\exp \left[-\frac{\left(X-\langle \hat{p}\rangle_t\right)^2}{2\Delta p_t^2}\right].
\end{eqnarray}
>From  Eqs. (\ref{pdfq})-(\ref{pdfp}) we see that the tomographic map
depends only on a single parameter $X$. This reduces the
dimensionality of  the problem with respect to the Wigner function,
that is a function of   both $p$ and $q$.
 The lines individuated by the choices $(\mu,\nu)=(1,0)$ and $(\mu,\nu)=(0,1)$ correspond to
tomograms depending on the time average and variance respectively of
position and momentum. In order to determine the latter quantities
we have to invert Eq. (\ref{pdfq}) and (\ref{pdfp}) for different
values of $X$, i.e. for a given number of points to measure
along a tomogram. Thus, our first goal is to determine the number
of tomograms required to measure the cumulants of our Gaussian
state.

\begin{figure}[h]
\begin{center}
\includegraphics[width=7 cm, height=6 cm]{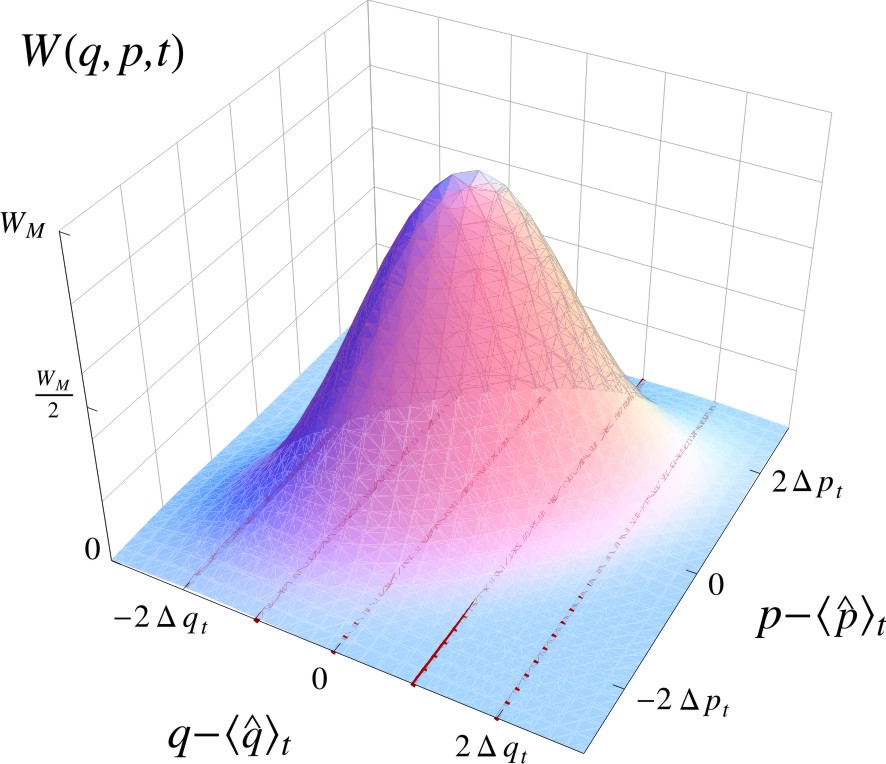}
\caption{\label{FigWigFun}\footnotesize{Wigner function, $W(q,p,t)$,
of Eq.~(\ref{wigner function gaussian}) and some straight lines
along the direction $\mu=1$ and $\nu=0$ on the plane $q p$. $\Delta
q_t^2\Delta p_t^2-\sigma(q,p)_t^2=0.64\Delta q_t^2\Delta
p_t^2$. $W_M= \frac{1}{2 \pi \sqrt{\Delta q_t^2\Delta
p_t^2-\sigma(q,p)_t^2}}$.}}
\end{center}
\end{figure}

\begin{figure}[h]
\begin{center}
\includegraphics[width=7 cm, height=5 cm]{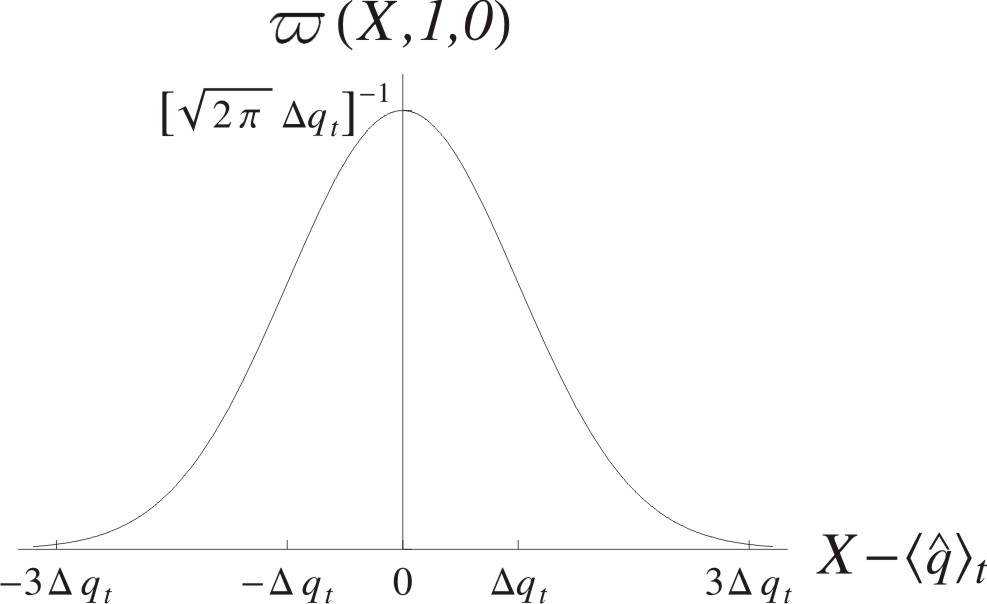}
\caption{\label{FigTomX}\footnotesize{Tomogram, $ \varpi (X,1,0)$,
of Eq.~(\ref{pdfq}) for the direction $\mu=1$ and $\nu=0$. }}
\end{center}
\end{figure}

\begin{figure}[h]
\begin{center}
\includegraphics[width=7 cm, height=5 cm]{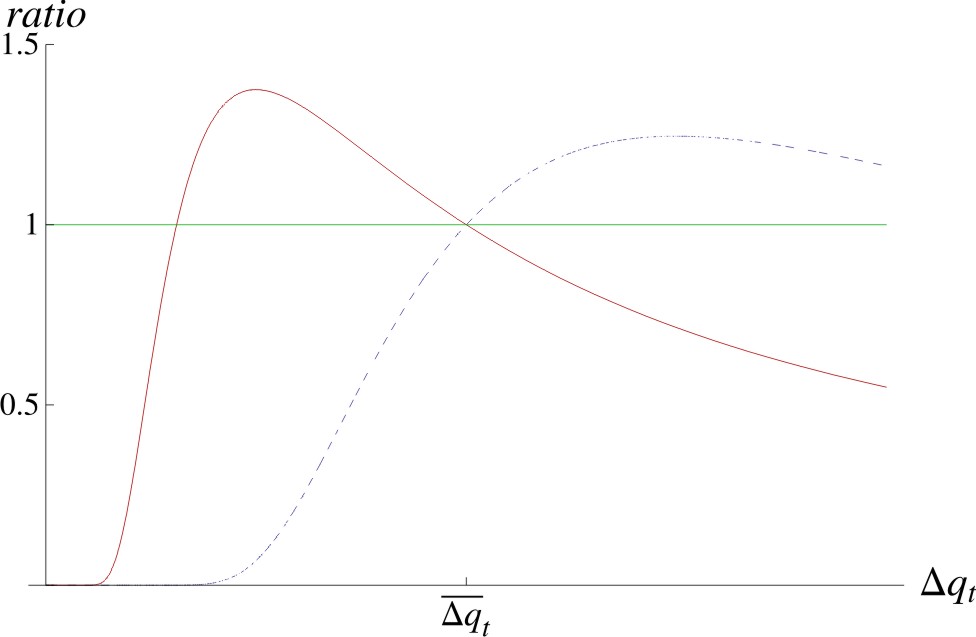}
\caption{\label{Figratio}\footnotesize{The ratio between right and
left side of Eq.~(\ref{eqtrascend}) for two different
values of $X$ with  $\mu=1$ and $\nu=0$ is plotted as a function of $\Delta q_t$. Values used:
$\langle \hat{q}\rangle_t=3$ and $X_1=4.5$ (continuous line) and
$X_2=2.5$ (dashed line). The values of $\varpi (X,1,0)$ are computed
using $\overline{\Delta q}_t=1$ to simulate what one would get
experimentally. Continuous line shows how Eq.~(\ref{eqtrascend})
with $X_1=4.5$ and $\varpi(4.5,1,0)$ can be satisfied (ratio=1) by
two values of  $\Delta q_t$. The comparison with a second case
with $X_2=2.5$ allows to determine which values of $\Delta q_t$ is
the right one. }}
\end{center}
\end{figure}

To answer this question, we first focus on the direction $\mu=1$,
$\nu=0$. In Fig.~\ref{FigWigFun} we plot the Wigner function of our
system at a generic time $t$ and some straight lines along the
considered direction. In Fig.~\ref{FigTomX} we plot the GSP
tomogram defined by Eq. (\ref{pdfq}). Inverting Eq.~(\ref{pdfq}),
we obtain:
\begin{equation}
\left(X-\langle \hat{q}\rangle_t\right)^2=2\Delta q_t^2\ln\frac{1}{\varpi(X,1,0)\Delta q_t\sqrt{2\pi}}\,.
\end{equation}
Using the value of the tomogram $\varpi(0,1,0)$ we can get $\langle \hat{q}\rangle_t$ as a function of $\Delta
q_t$:
\begin{equation}\label{media di q from tomigram}
    \langle \hat{q}\rangle_t=\pm\Delta q_t\sqrt{2\ln\frac{1}{\varpi(0,1,0)\Delta q_t\sqrt{2\pi}}}\,.
\end{equation}
If we know the sign of $\langle \hat{q}\rangle_t$ then we
need only the value of the tomogram $\varpi(0,1,0)$ to get
$\langle \hat{q}\rangle_t$, otherwise we need another point.
Using Eq.~(\ref{media di q from tomigram}),  Eq.~(\ref{pdfq})
becomes an equation for $\Delta q_t$ only, and it can be rewritten
as
\begin{eqnarray} \label{eqtrascend} 2\Delta
q_t^2&&\ln\frac{1}{\varpi(X,1,0)\Delta q_t\sqrt{2\pi}}\nonumber
\\&&=\left(X\mp\Delta q_t\sqrt{2\ln\frac{1}{\varpi(0,1,0)\Delta
q_t\sqrt{2\pi}}}\right)^2. \nonumber \\
\end{eqnarray}
This equation is trascendental, therefore we will solve it
numerically. We can graphically note in Fig.~\ref{Figratio} that for each $X$ and
corresponding $\varpi(X,1,0)$ there may be two values of $\Delta
q_t$ satisfying the previous equation. In order to identify one of
the two solutions, it is enough to consider two points,
$\left\{(X_1,\varpi(X_1,1,0))\right\}$ and
$\left\{(X_2,\varpi(X_2,1,0))\right\}$, and to choose the common
solution for the variance. This  is made clear by
Fig.~\ref{Figratio}, where the ratio between right and left side of
Eq.~(\ref{eqtrascend}) for two different values of $X$ is plotted.
The common solution (i.e. when both ratios are equal to 1) is
labeled $\overline{\Delta q}_t$.

As a consequence, whether we know or not the sign of the average $\langle
q\rangle_t$, we need three or four points to determine
$\langle \hat{q}\rangle_t$ and $\Delta q_t$ in Eq. (\ref{pdfq}).
Analogously, we need other three or four points for
$\langle \hat{p}\rangle_t$ and $\Delta p_t$ in Eq.~(\ref{pdfp}).

Let us now compute the covariance
$\sigma(q,p)_t$. To this purpose, we consider the tomogram:
\begin{multline}
\varpi\left(X,\frac{1}{\sqrt{2}},\frac{1}{\sqrt{2}}\right)=\frac{1
}{\sqrt{\pi}\sqrt{\Delta q_t^2+\Delta p_t^2 +
2\sigma(q,p)_t}}\\\exp
\left[-\frac{\left(X-\frac{\langle \hat{q}\rangle_t + \langle
\hat{p}\rangle_t}{\sqrt{2}} \right)^2}{\Delta q_t^2+\Delta
p_t^2 + 2\sigma(q,p)_t}\right].
\end{multline}
This is  a Gaussian whose average value is already determined.  Indeed,
according to the previous steps, we need two more points of
this tomograms to determine the spread $(\Delta q_t^2+\Delta
p_t^2)/2 + \sigma(q,p)_t$, from which we can retrieve
$\sigma(q,p)_t$.

Hence, we have shown that by means of eight or at most ten
points belonging to three tomograms, the first and second
order momenta of a Gaussian state can be measured at an arbitrary
time $t$.   One can then use these measured cumulants in order to
infer the master equation parameters describing the system under
investigation. We note also that we can reasonably  infer that the
number of tomograms needed to reconstruct the system density
operator is minimized by employing Gaussian wave packets as a
probe. Indeed these states have minimum uncertainty, and are the
only states having positive Wigner function \cite{Folland1989}.

\section{Conclusions}\label{par:Summary and Conclusions}

In this paper we have proposed an approach to the study of open
quantum systems based on quantum symplectic tomography.

In many contexts the reduced dynamics of a system coupled with its
environment is modeled by  phenomenological master equations with
some general features, but with unknown parameters. Hence, it would
be highly appealing  to find a way to assign some values to these
parameters. We have tackled this problem for a wide class of
Markovian master equations, which are the Gaussian-shape-preserving
ones. We have proved that it is possible to retrieve their unknown
parameters by performing a limited number (ten at maximum) of time-independent
measurements using Gaussian wave packets as a probe.

This result leads to some interesting applications. Once retrieved the unknown master equation coefficients,
it is possible to compute the dynamical evolution of
any physical quantity whose analytical expression is known.
The indirect-measurement scheme we propose
could be then employed to make predictions on system loss of coherence due to the external
environment. In order to perform this kind of analysis one can
consider some quantities such as the spread and the coherence length
in both position and momentum \cite{Barnett2001}, provided their
analytical expressions are available for an arbitrary time $t$ (e.g.
see Ref.\cite{bellomo2007a}).
Working
in the coherent state representation, the
evolution of the system of interest from an arbitrary initial state can be in principle predicted. Therefore, it is  possible to perform the proposed indirect analysis  of the decoherence processes.
For example, if we consider an initial
Schr\"{o}dinger-cat state, highly interesting
due to its potentially long-range coherence properties and its
extreme sensitivity  to environmental decoherence \cite{Brune1997}, we can re-write it as a combination of four Gaussian functions. Therefore, due to the linearity of the master equation, it can be possible to derive analytically
the state evolution and to analyze its loss of coherence by means of the procedure we propose.
\section{Acknowledgements}
 We warmly thank Dr. P. Facchi, Prof. G. Marmo and Prof. S.
Pascazio for many interesting and useful discussions. In particular
we thank Prof. G. Marmo for his invitation at the University of Naples "Federico II" which gave us the chance of starting this work.

\appendix

\section{Alternative procedure}\label{Par:Alternative procedure}

Here we propose an alternative time-dependent procedure to compute  the
second cumulants of a Gaussian state, by means of tomograms, given the knowledge
of the first cumulants time evolution. To this purpose
we need to consider the following three tomograms:

\begin{eqnarray}
  \varpi_1&=&\varpi (\langle \hat{p}\rangle_t,0,1) =
    \frac{1 }{\sqrt{2\pi}\Delta p_t} \nonumber\\
\varpi_2&=&  \varpi (\langle \hat{q}\rangle_t,1,0) =  \frac{1 }{\sqrt{2\pi}\Delta q_t} \nonumber  \\
  \varpi_3&=&\varpi \left(\frac{\langle \hat{p}\rangle_t+\langle \hat{q}\rangle_t}{\sqrt{2}},\frac{1}
  {\sqrt{2}},\frac{1}{\sqrt{2}}\right)\nonumber\\&=&  \frac{1 }{\sqrt{2\pi}\sqrt{\Delta q_t^2/2+\Delta p_t^2 /2 + \sigma(q,p)_t }}.
\end{eqnarray}
Inverting the previous equations one can infer $\Delta q_t$, $\Delta
p_t$ and $\sigma(q,p)_t$ from the knowledge of $\varpi_1$,
$\varpi_2$ and $\varpi_3$.
However, this procedure presents two drawbacks. In fact,  the evolved averaged
values $\langle \hat{q}\rangle_t$ and $\langle \hat{p}\rangle_t$
are required and  we need tomograms evaluated  on time-dependent
variables. These problems  do not arise in the  time-independent procedure,
 based only on tomograms for which no a priori knowledge on the Gaussian state is required.
Nevertheless, in  this alternative  time-dependent scheme only three tomograms are required.

\end{document}